\documentclass[10pt,journal,epsfig]{IEEEtran}

\usepackage[dvips]{graphicx}
\usepackage{graphicx}
\usepackage{amssymb}
\usepackage{cite}
\usepackage{subfigure}
\usepackage{amsmath}

\begin{document}

\title{Super-Resolution Compressed Sensing: An Iterative Reweighted Algorithm for Joint Parameter Learning
and Sparse Signal Recovery}

\author{Jun Fang, Jing Li, Yanning Shen, Hongbin Li,~\IEEEmembership{Senior
Member,~IEEE}, and Shaoqian Li
\thanks{Jun Fang, Jing Li, Yanning Shen and Shaoqian Li are with the National Key Laboratory on Communications,
University of Electronic Science and Technology of China, Chengdu
611731, China, Emails: JunFang@uestc.edu.cn,
201221260204@std.uestc.edu.cn, 201121260110@std.uestc.edu.cn,
lsq@uestc.edu.cn}
\thanks{Hongbin Li is
with the Department of Electrical and Computer Engineering,
Stevens Institute of Technology, Hoboken, NJ 07030, USA, E-mail:
Hongbin.Li@stevens.edu}
\thanks{This work was supported in part by the National Science
Foundation of China under Grant 61172114, and the National Science
Foundation under Grant ECCS-0901066. }}

%\thanks{Huiping Duan is with the School of Electronic Engineering,
%University of Electronic Science and Technology of China, Chengdu
%611731, China, Email: huipingduan@uestc.edu.cn}

\maketitle

%Conventional compressed sensing theory assumes signals have sparse
%representations in a known, finite dictionary. Nevertheless

%based on which a ``presumed dictionary'' is constructed for sparse
%signal recovery.

%since the true parameters do not necessarily lie on the
%discretized grid. This error, also referred to as grid mismatch,

%Unlike the conventional compressed sensing approach,

%The proposed algorithm does not need to specify presumed
%dictionary in advance. Instead, ,

%The algorithm renders an accurate estimate of the unknown
%parameters as well as the sparse signals.

\begin{abstract}
In many practical applications such as direction-of-arrival (DOA)
estimation and line spectral estimation, the sparsifying
dictionary is usually characterized by a set of unknown parameters
in a continuous domain. To apply the conventional compressed
sensing to such applications, the continuous parameter space has
to be discretized to a finite set of grid points. Discretization,
however, incurs errors and leads to deteriorated recovery
performance. To address this issue, we propose an iterative
reweighted method which jointly estimates the unknown parameters
and the sparse signals. Specifically, the proposed algorithm is
developed by iteratively decreasing a surrogate function
majorizing a given objective function, which results in a gradual
and interweaved iterative process to refine the unknown parameters
and the sparse signal. Numerical results show that the algorithm
provides superior performance in resolving closely-spaced
frequency components.
\end{abstract}

%effectiveness and superiority over other existing methods.


%The principal idea behind the sparse signal recovery technique can
%be stated as seeking for the sparsest signal based on the linear
%measurements
%\begin{align}
%\min_{\boldsymbol{x}} & \|\boldsymbol{x}\|_0\nonumber\\
%\text{s.t.} & \phantom{0}
%\boldsymbol{A}\boldsymbol{x}=\boldsymbol{y}\label{opt0}
%\end{align}
%where the notation $\|\cdot\|_0$ refers to the number of non-zero
%coefficients of the vector. This optimization problem, however, is
%NP-hard and has a computational complexity growing exponentially
%with the signal dimension $n$. Thus, alternative
%sparsity-promoting functionals which are more computationally
%efficient in finding the sparse solution are desirable.

%problem of estimating the frequency components of a mixture of
%complex sinusoids

%The same is true for direction-of-arrival (DOA) estimation and
%source localization in sensor networks, where the true directions
%or locations of the sources may not be consistent with the
%presumed grid \cite{YangXie11,FengValaee09}. Overall, in these
%applications, the sparsifying dictionary is usually characterized
%by a set of unknown parameters in a continuous domain. In order to
%apply the compressed sensing technique to such applications, the
%continuous parameter space has to be discretized to a finite set
%of grid points, based on which a presumed dictionary is
%constructed for sparse signal recovery. Discretization, however,
%inevitably incurs errors since the true parameters do not
%necessarily lie on the discretized grid.

\section{Introduction}
The compressed sensing technique finds a variety of applications
in practice as many natural signals admit a sparse or an
approximate sparse representation in a certain basis.
Nevertheless, the accurate reconstruction of the sparse signal
relies on the knowledge of the sparsifying dictionary. While in
many applications, it is often impractical to preset a dictionary
that can sparsely represent the signal. For example, for the line
spectral estimation problem, using a preset discrete Fourier
transform (DFT) matrix suffers from a considerable performance
degradation because the true frequency components may not lie on
the pre-specified frequency grid
\cite{CandesGranda12,TangBhaskar12}. This discretization error is
also referred to as the grid mismatch.

%Finer grids can certainly be used to reduce the grid mismatch and
%improve the reconstruction accuracy. Nevertheless, recovery
%algorithms become numerically instable when very fine discretized
%grids are employed.

%leads to deteriorated performance or even failure in recovering
%the sparse signals.

%Specifically, in \cite{ChiScharf11}, the problem was addressed in
%a general framework of ``basis mismatch'' where there is a
%perturbation (caused by grid discretization, calibration errors or
%other factors) between the presumed and the actual dictionaries,
%and the impact of the basis mismatch on the reconstruction error
%was analyzed.

%Bayesian methods \cite{YangXie13,HuZhou13} were developed to
%estimate the sparse signal along with the unknown parameters
%associated with the structured matrix.

The grid mismatch problem has attracted a lot of attention over
the past few years, e.g.
\cite{ChiScharf11,YangXie13,HuZhou13,FannjiangLiao12,DuarteBaraniuk13,CandesGranda12,TangBhaskar12,HuShi12}.
Specifically, in \cite{YangXie13,HuZhou13}, to deal with the grid
mismatch, the true dictionary is approximated as a summation of a
presumed dictionary and a structured parameterized matrix via the
Taylor expansion. The recovery performance of this method,
however, depends on the accuracy of the Taylor expansion in
approximating the true dictionary. The grid mismatch problem was
also examined in \cite{FannjiangLiao12,DuarteBaraniuk13}, where a
highly coherent dictionary (very fine grids) is used to mitigate
the discretization error, and the technique of band exclusion
(coherence-inhibiting) was proposed for sparse signal recovery.
Besides these efforts, another line of work
\cite{CandesGranda12,TangBhaskar12,HuShi12} studied the problem of
grid mismatch in an undirect but more fundamental way: they
circumvent the discretization issue by working directly on the
continuous parameter space (this approach is also referred to as
super-resolution techniques). In
\cite{CandesGranda12,TangBhaskar12}, an atomic norm-minimization
and a total variation norm-minimization approaches were proposed
to handle the infinite dictionary with continuous atoms.
Nevertheless, finding a solution to the total variation or atomic
norm problem is challenging. Although the total variation norm
problem can be cast into a convex semidefinite program
optimization for the complex sinusoid mixture problem, it still
remains unclear how this reformulation generalizes to other
scenarios. In \cite{HuShi12}, by treating the sparse signal as
hidden variables, a Bayesian approach was proposed to jointly
iteratively refine the dictionary, and is shown able to achieve
super-resolution accuracy.

In this paper, we propose an iterative reweighted method for joint
parameter learning and sparse signal recovery. The algorithm is
developed by iteratively decreasing a surrogate function that
majorizes the original objective function. Our experiments show
that our proposed algorithm achieves a significant performance
improvement as compared with existing methods in distinguishing
and recovering complex sinusoids whose frequencies are very
closely separated.

\section{Problem Formulation}
In many practical applications such as direction-of-arrival (DOA)
estimation and line spectral estimation, the sparsifying
dictionary is usually characterized by a set of unknown parameters
in a continuous domain. For example, consider the line spectral
estimation problem where the observed signal is a summation of a
number of complex sinusoids:
\begin{align}
y_m=\sum_{k=1}^K \alpha_k e^{-j\omega_k m} \qquad m=1,\ldots, M
\label{data-model}
\end{align}
where $\omega_k\in [0, 2\pi)$ and $\alpha_k$ denote the frequency
and the complex amplitude of the $k$-th component, respectively.
Define
$\boldsymbol{a}(\theta)\triangleq[e^{-j\omega}\phantom{0}e^{-j2\omega}\phantom{0}\ldots\phantom{0}e^{-jM\omega}
]^T$, the model (\ref{data-model}) can be rewritten in a
vector-matrix form as
\begin{align}
\boldsymbol{y}=\boldsymbol{A}(\boldsymbol{\omega})\boldsymbol{\alpha}
\end{align}
where $\boldsymbol{y}\triangleq
[y_1\phantom{0}\ldots\phantom{0}y_M]^T$,
$\boldsymbol{\alpha}\triangleq
[\alpha_1\phantom{0}\ldots\phantom{0}\alpha_K]^T$, and
$\boldsymbol{A}(\boldsymbol{\omega})\triangleq
[\boldsymbol{a}(\omega_1)\phantom{0}\ldots\phantom{0}\boldsymbol{a}(\omega_K)]$.
We see that the dictionary $\boldsymbol{A}(\boldsymbol{\omega})$
is characterized by a number of unknown parameters $\{\omega_k\}$
which needs to be estimated along with the unknown complex
amplitudes $\{\alpha_k\}$. To deal with this problem, conventional
compressed sensing techniques discretize the continuous parameter
space into a finite set of grid points, assuming that the unknown
frequency components $\{\omega_k\}$ lie on the discretized grid.
Estimating $\{\omega_k\}$ and $\{\alpha_k\}$ can then be
formulated as a sparse signal recovery problem
$\boldsymbol{y}=\boldsymbol{Ax}$, where
$\boldsymbol{A}\in\mathbb{C}^{M\times N}$ ($M\ll N$) is an
overcomplete dictionary constructed based on the discretized grid
points. Discretization, however, inevitably incurs errors since
the true parameters do not necessarily lie on the discretized
grid. This error, also referred to as the grid mismatch, leads to
deteriorated performance or even failure in recovering the sparse
signals.

%unlike the conventional compressed sensing theory which assumes a
%presumed dictionary,

To circumvent this issue, we treat the overcomplete dictionary as
an unknown parameterized matrix
$\boldsymbol{A}(\boldsymbol{\theta})\triangleq
[\boldsymbol{a}(\theta_1)\phantom{0}\ldots\phantom{0}\boldsymbol{a}(\theta_N)]$,
with each atom $\boldsymbol{a}(\theta_n)$ determined by an unknown
frequency parameter $\theta_n$. Estimating $\{\omega_k\}$ and
$\{\alpha_k\}$ can still be formulated as a sparse signal recovery
problem. Nevertheless, in this framework, the frequency parameters
$\boldsymbol{\theta}\triangleq \{\theta_n\}_{n=1}^N$ need to be
optimized along with the sparse signal such that the parametric
dictionary will approach the true sparsifying dictionary.
Specifically, the problem of joint parameter learning and sparse
signal recovery can be presented as follows: we search for a set
of unknown parameters $\{\theta_n\}_{n=1}^N$ with which the
observed signal $\boldsymbol{y}$ can be represented by as few
atoms as possible. Such a problem can be readily formulated as
\begin{align}
\min_{\boldsymbol{z},\boldsymbol{\theta}}\quad &
\|\boldsymbol{z}\|_0 \nonumber\\
\text{s.t.} \quad
&\boldsymbol{y}=\boldsymbol{A}(\boldsymbol{\theta})\boldsymbol{z}
\label{opt-1}
\end{align}
where $\|\boldsymbol{z}\|_0$ stands for the number of the nonzero
components of $\boldsymbol{z}$. The optimization (\ref{opt-1}),
however, is an NP-hard problem that has computational complexity
growing exponentially with the signal dimension $N$. Thus,
alternative sparsity-promoting functionals which are more
computationally efficient in finding the sparse solution are
desirable. In this paper, we consider the use of the log-sum
sparsity-encouraging functional for sparse signal recovery.
Log-sum penalty function was originally introduced in
\cite{CoifmanWickerhauser92} for basis selection and has been
extensively used for sparse signal recovery, e.g.
\cite{GorodnitskyRao97,CandesWakin08,ChartrandYin08}. It was
proved theoretically \cite{ShenFang13} and shown in a series of
experiments \cite{CandesWakin08} that log-sum based methods
present uniform superiority over the conventional $\ell_1$-type
methods. Replacing the $\ell_0$-norm in (\ref{opt-1}) with the
log-sum functional leads to
\begin{align}
\min_{\boldsymbol{z},\boldsymbol{\theta}}\quad &
L(\boldsymbol{z})=\sum_{i=1}^N \log
(|z_i|^2+\epsilon)\quad \nonumber\\
\text{s.t.} \quad
&\boldsymbol{y}=\boldsymbol{A}(\boldsymbol{\theta})\boldsymbol{z}
\label{opt-2}
\end{align}
where $z_i$ denotes the $i$th component of the vector
$\boldsymbol{z}$, and $\epsilon>0$ is a positive parameter to
ensure that the function is well-defined. Note that the above
optimization (\ref{opt-2}) can be formulated as an unconstrained
optimization problem by removing the constraint and adding a
penalty term,
$\lambda\|\boldsymbol{y}-\boldsymbol{A}(\boldsymbol{\theta})\boldsymbol{z}\|_2^2$,
to the objective functional. A two-stage iterative algorithm
\cite{AtaeeZayyani10} can then be applied: given an estimate of
$\boldsymbol{\theta}$, the sparse signal $\boldsymbol{z}$ is
recovered using conventional compressive sensing techniques; and
estimate $\boldsymbol{\theta}$ based on the estimated
$\boldsymbol{z}$. This scheme, however, is computationally
expensive because it requires to solve the sparse signal recovery
problem every iteration. The trade-off parameter $\lambda$ is also
difficult to determine due to the non-convexity of the objective
function. In addition, the two-stage algorithm is very likely to
be trapped in undesirable local minima, possibly because the
estimated signal, instead of optimized in a gradual manner,
undergoes an abrupt change from one iteration to another and thus
easily deviates from the correct basin of attraction. In the
following, we develop an iterative reweighted algorithm which less
likely suffers from the local convergence issue.

\section{Proposed Algorithm}
The proposed algorithm is developed based on a bounded
optimization approach, also known as the majorization-minimization
approach \cite{LangeHunter00,CandesWakin08}. The idea is to
iteratively minimize a simple surrogate function majorizing a
given objective function. A surrogate function, usually written as
$Q(\boldsymbol{z}|\hat{\boldsymbol{z}}^{(t)})$, is an upper bound
for the objective function $L(\boldsymbol{z})$. Precisely, we have
\begin{align}
Q(\boldsymbol{z}|\boldsymbol{\hat{z}}^{(t)})-L(\boldsymbol{z})\geq
0
\end{align}
with the equality attained when
$\boldsymbol{z}=\boldsymbol{\hat{z}}^{(t)}$. We will show that
through iteratively decreasing (not necessarily minimizing) the
surrogate function, the iterative process yields a non-increasing
objective function value and eventually converges to a stationary
point of $L(\boldsymbol{x})$.

%so that the minimization of the surrogate function admits an
%analytical solution, or at least is a well-behaved numerical
%problem

We first discuss how to find a surrogate function for the
objective function defined in (\ref{opt-2}). Ideally, we hope that
the surrogate function is differentiable and convex. An
appropriate choice of such a surrogate function has a quadratic
form and is given by
\begin{align}
Q(\boldsymbol{z}|\boldsymbol{\hat{z}}^{(t)})\triangleq\sum_{i=1}^N
\bigg(\frac{|z_i|^2+\epsilon}{|\hat{z}_i^{(t)}|^2+\epsilon}+\log(|\hat{z}_i^{(t)}|^2+\epsilon)-1
\bigg) \label{surrogate-function}
\end{align}
It can be readily verified that
\begin{align}
Q(\boldsymbol{z}|\boldsymbol{\hat{z}}^{(t)})-L(\boldsymbol{z})
\geq& 0
\end{align}
where the inequality becomes equality when
$\boldsymbol{z}=\boldsymbol{\hat{z}}^{(t)}$. The convex quadratic
function $Q(\boldsymbol{z}|\boldsymbol{\hat{z}}^{(t)})$ is
therefore a surrogate function for the log-sum
sparsity-encouraging functional. Replacing the log-sum functional
in (\ref{opt-2}) with (\ref{surrogate-function}), we arrive at the
following optimization
\begin{align}
\min_{\boldsymbol{z},\boldsymbol{\theta}}\quad &
\boldsymbol{z}^H\boldsymbol{D}^{(t)}\boldsymbol{z} \nonumber\\
\text{s.t.} \quad
&\boldsymbol{y}=\boldsymbol{A}(\boldsymbol{\theta})\boldsymbol{z}
\label{opt-4}
\end{align}
where $[\cdot]^{H}$ denotes the conjugate transpose, and
$\boldsymbol{D}^{(t)}$ is a diagonal matrix given as
\begin{align}
\boldsymbol{D}^{(t)}\triangleq\text{diag}
\bigg\{\frac{1}{|\hat{z}_1^{(t)}|^2+\epsilon},\ldots,\frac{1}{|\hat{z}_N^{(t)}|^2+\epsilon}\bigg\}
\nonumber
\end{align}
Given $\boldsymbol{\theta}$ fixed, the optimal $\boldsymbol{z}$ of
(\ref{opt-4}) can be obtained by resorting to the Lagrangian
multiplier method and given as
\begin{align}
\boldsymbol{z}&=(\boldsymbol{D}^{(t)})^{-1}\boldsymbol{A}^H(\boldsymbol{\theta})\left(\boldsymbol{A}(\boldsymbol{\theta})
(\boldsymbol{D}^{(t)})^{-1}\boldsymbol{A}^H(\boldsymbol{\theta})\right)^{-1}\boldsymbol{y}
\label{eq3}
\end{align}
Substituting (\ref{eq3}) back into (\ref{opt-4}), the optimization
simply becomes searching for the unknown parameter
$\boldsymbol{\theta}$:
\begin{align}
\min_{\boldsymbol{\theta}}\phantom{0}\boldsymbol{y}^{H}\left(\boldsymbol{A}(\boldsymbol{\theta})
(\boldsymbol{D}^{(t)})^{-1}\boldsymbol{A}^H(\boldsymbol{\theta})\right)^{-1}\boldsymbol{y}
\label{opt-5}
\end{align}
An analytical solution of the above optimization (\ref{opt-5}) is
difficult to obtain. Nevertheless, in our algorithm, we only need
to search for a new estimate $\boldsymbol{\hat{\theta}}^{(t+1)}$
such that the following inequality holds valid
\begin{align}
&\boldsymbol{y}^{H}\left(\boldsymbol{A}(\boldsymbol{\hat{\theta}}^{(t+1)})
(\boldsymbol{D}^{(t)})^{-1}\boldsymbol{A}^H(\boldsymbol{\hat{\theta}}^{(t+1)})\right)^{-1}\boldsymbol{y}
\nonumber\\
&\leq
(\boldsymbol{\hat{z}}^{(t)})^H\boldsymbol{D}^{(t)}\boldsymbol{\hat{z}}^{(t)}
\label{eq4}
\end{align}
Such an estimate can be found by using the gradient descent
method. Note that since the optimizations (\ref{opt-5}) and
(\ref{opt-4}) attain the same minimum objective function value, we
can always find an estimate $\boldsymbol{\hat{\theta}}^{(t+1)}$ to
meet (\ref{eq4}). In fact, our experiments suggest that finding
such an estimate is much easier than searching for a local or
global minimum of the optimization (\ref{opt-5}).

%Also, due to the non-convexity of the objective function,
%convergence to the global minimum is not guaranteed by any
%gradient-based search methods.
%we do not need to find the global minimum of (\ref{opt-5}).
%Instead,

Given $\boldsymbol{\hat{\theta}}^{(t+1)}$,
$\boldsymbol{\hat{z}}^{(t+1)}$ can be obtained via (\ref{eq3}),
with $\boldsymbol{\theta}$ replaced by
$\boldsymbol{\hat{\theta}}^{(t+1)}$, i.e.
\begin{align}
\boldsymbol{\hat{z}}^{(t+1)}=(\boldsymbol{D}^{(t)})^{-1}
&\boldsymbol{A}^H(\boldsymbol{\hat{\theta}}^{(t+1)}) \nonumber\\
\times &\left(\boldsymbol{A}(\boldsymbol{\hat{\theta}}^{(t+1)})
(\boldsymbol{D}^{(t)})^{-1}\boldsymbol{A}^H(\boldsymbol{\hat{\theta}}^{(t+1)})\right)^{-1}\boldsymbol{y}
\label{eq7}
\end{align}
In the following, we will show that the new obtained estimate
$\boldsymbol{\hat{z}}^{(t+1)}$ results in a non-increasing
objective function value, that is,
$L(\boldsymbol{\hat{z}}^{(t+1)})\leq
L(\boldsymbol{\hat{z}}^{(t)})$. Firstly, we have
\begin{align}
Q(\boldsymbol{\hat{z}}^{(t+1)}|\boldsymbol{\hat{z}}^{(t)})=&
(\boldsymbol{\hat{z}}^{(t+1)})^H\boldsymbol{D}^{(t)}\boldsymbol{\hat{z}}^{(t+1)}
\nonumber\\
=&\boldsymbol{y}^{H}\left(\boldsymbol{A}(\boldsymbol{\hat{\theta}}^{(t+1)})
(\boldsymbol{D}^{(t)})^{-1}\boldsymbol{A}(\boldsymbol{\hat{\theta}}^{(t+1)})^H\right)^{-1}\boldsymbol{y}
\nonumber\\
\stackrel{(a)}{\leq}&
(\boldsymbol{\hat{z}}^{(t)})^H\boldsymbol{D}^{(t)}\boldsymbol{\hat{z}}^{(t)}=
Q(\boldsymbol{\hat{z}}^{(t)}|\boldsymbol{\hat{z}}^{(t)})
\label{eq5}
\end{align}
where $(a)$ comes from the inequality (\ref{eq4}). Based on
(\ref{eq5}), we reach the following
\begin{align}
L(\boldsymbol{\hat{z}}^{(t+1)})=&L(\boldsymbol{\hat{z}}^{(t+1)})-Q(\boldsymbol{\hat{z}}^{(t+1)}|\boldsymbol{\hat{z}}^{(t)})
+Q(\boldsymbol{\hat{z}}^{(t+1)}|\boldsymbol{\hat{z}}^{(t)})
\nonumber\\
\leq &
L(\boldsymbol{\hat{z}}^{(t)})-Q(\boldsymbol{\hat{z}}^{(t)}|\boldsymbol{\hat{z}}^{(t)})
+Q(\boldsymbol{\hat{z}}^{(t+1)}|\boldsymbol{\hat{z}}^{(t)})
\nonumber\\
\leq &
L(\boldsymbol{\hat{z}}^{(t)})-Q(\boldsymbol{\hat{z}}^{(t)}|\boldsymbol{\hat{z}}^{(t)})
+Q(\boldsymbol{\hat{z}}^{(t)}|\boldsymbol{\hat{z}}^{(t)})
\nonumber\\
=& L(\boldsymbol{\hat{z}}^{(t)}) \label{eq6}
\end{align}
where the first inequality follows from the fact that
$Q(\boldsymbol{z}|\boldsymbol{\hat{z}}^{(t)})-L(\boldsymbol{z})$
attains its minimum when
$\boldsymbol{z}=\boldsymbol{\hat{z}}^{(t)}$, the second inequality
follows from (\ref{eq5}). We see that through iteratively
decreasing (not necessarily minimizing) the surrogate function,
the objective function $L(\boldsymbol{z})$ is guaranteed to be
non-increasing at each iteration.

For clarity, we summarize our algorithm as follows.
\begin{enumerate}
\item Given an initialization
$\boldsymbol{\hat{z}}^{(0)}$.
\item At iteration $t=0,1,\ldots$: Based on the estimate $\boldsymbol{\hat{z}}^{(t)}$,
construct the surrogate function as depicted in
(\ref{surrogate-function}). Search for a new estimate of the
unknown parameter vector, denoted as
$\boldsymbol{\hat{\theta}}^{(t+1)}$, by using the gradient descent
method such that the inequality (\ref{eq4}) is satisfied. Compute
a new estimate of the sparse signal, denoted as
$\boldsymbol{\hat{z}}^{(t+1)}$, via (\ref{eq7}).
\item Go to Step 2 if
$\|\boldsymbol{\hat{z}}^{(t+1)}-\boldsymbol{\hat{z}}^{(t)}\|_2>\varepsilon$,
where $\varepsilon$ is a prescribed tolerance value; otherwise
stop.
\end{enumerate}

The second step of the proposed algorithm involves searching for a
new estimate of the unknown parameter vector to meet the condition
(\ref{eq4}). As mentioned earlier, this can be accomplished via a
gradient-based search algorithm. Define
\begin{align}
f(\boldsymbol{\theta})\triangleq&
\boldsymbol{y}^{H}(\boldsymbol{A}(\boldsymbol{\theta})
(\boldsymbol{D}^{(t)})^{-1}\boldsymbol{A}^H(\boldsymbol{\theta}))^{-1}\boldsymbol{y}
\nonumber\\
\boldsymbol{X}\triangleq &\boldsymbol{A}(\boldsymbol{\theta})
(\boldsymbol{D}^{(t)})^{-1}\boldsymbol{A}^H(\boldsymbol{\theta})
\nonumber
\end{align}
Using the chain rule, the first derivative of
$f(\boldsymbol{\theta})$ with respect to $\theta_i,\forall i$ can
be computed as
\begin{align}
\frac{\partial
f(\boldsymbol{\theta})}{\theta_i}=\text{tr}\left\{\bigg(\frac{\partial
f(\boldsymbol{\theta})}{\partial
\boldsymbol{X}}\bigg)^T\frac{\partial \boldsymbol{X}}{\partial
\theta_i}\right\}+\text{tr}\left\{\bigg(\frac{\partial
f(\boldsymbol{\theta})}{\partial
\boldsymbol{X}^{\ast}}\bigg)^T\frac{\partial
\boldsymbol{X}^{\ast}}{\partial \theta_i}\right\}
\end{align}
where $\boldsymbol{X}^{\ast}$ denotes the conjugate of the complex
matrix $\boldsymbol{X}$, and
\begin{align}
\frac{\partial f(\boldsymbol{\theta})}{\partial
\boldsymbol{X}}=&\frac{\partial }{\partial
\boldsymbol{X}}\text{tr}\left\{\boldsymbol{y}\boldsymbol{y}^{H}\boldsymbol{X}^{-1}\right\}
\nonumber\\
=&-\left(\boldsymbol{X}^{-1}\boldsymbol{y}\boldsymbol{y}^{H}\boldsymbol{X}^{-1}\right)^T
\nonumber
\end{align}
\begin{align}
\frac{\partial f(\boldsymbol{\theta})}{\partial
\boldsymbol{X}^{\ast}}=&\frac{\partial }{\partial
\boldsymbol{X}^{\ast}}\text{tr}\left\{\boldsymbol{y}\boldsymbol{y}^{H}\boldsymbol{X}^{-1}\right\}=
\boldsymbol{0} \nonumber
\end{align}
\begin{align}
\frac{\partial \boldsymbol{X}}{\partial\theta_i}=&\frac{\partial
}{\partial\theta_i}\boldsymbol{A}(\boldsymbol{\theta})
(\boldsymbol{D}^{(t)})^{-1}\boldsymbol{A}^H(\boldsymbol{\theta})
\nonumber\\
=&\frac{\partial\boldsymbol{A}(\boldsymbol{\theta})
}{\partial\theta_i}(\boldsymbol{D}^{(t)})^{-1}\boldsymbol{A}^H(\boldsymbol{\theta})
+\boldsymbol{A}(\boldsymbol{\theta})
(\boldsymbol{D}^{(t)})^{-1}\frac{\partial\boldsymbol{A}^H(\boldsymbol{\theta})
}{\partial\theta_i} \nonumber
\end{align}
The current estimate $\boldsymbol{\hat{\theta}}^{(t)}$ can be used
as an initialization point to search for the new estimate
$\boldsymbol{\hat{\theta}}^{(t+1)}$. Our experiments suggest that
a new estimate which satisfies (\ref{eq4}) can usually be obtained
within only a few iterations. When the iterations achieve a steady
state, the estimates of $\{\theta_i\}$ can be refined in a
sequential manner to help achieve a better reconstruction
accuracy, but only those parameters whose coefficients are
relatively large are required to be updated every iteration.

%The proposed algorithm can be considered as consisting of two
%alternating steps. First, we estimate the signal $\boldsymbol{x}$
%and the unknown parameters $\boldsymbol{\theta}$ through
%decreasing the current surrogate function
%$Q(\boldsymbol{x}|\boldsymbol{\hat{x}}^{(t)})$. Second, based on
%the estimate of $\boldsymbol{x}$, we update the weights of the
%surrogate function and construct a new surrogate function. This
%iterative reweighted process is similar to the conventional
%iterative reweighted algorithms discussed in
%\cite{ChartrandYin08,DaubechiesDevore10}. Nevertheless, our
%proposed algorithm needs to estimate the sparse signal as well as
%the unknown parameters.

We see that in our algorithm, the unknown parameters and the
signal are refined in a gradual and interweaved manner. This
interweaved and gradual refinement enables the algorithm, with a
high probability, comes to a reasonably nearby point during the
first few iterations, and eventually converges to the correct
basin of attraction. In addition, like \cite{ChartrandYin08}, we
can improve the ability of avoiding undesirable local minima by
using a monotonically decreasing sequence $\{\epsilon^{(t)}\}$,
instead of a constant $\epsilon$, in updating the weighting
parameters in (\ref{surrogate-function}). For example, at the
beginning, $\epsilon^{(0)}$ can be set to a relatively large
value, say 1, in order to provide a stable coefficient estimate.
We then gradually reduce the value of $\epsilon^{(t)}$ in the
subsequent iterations until $\epsilon^{(t)}$ attains a prescribed
value, say, $10^{-8}$.

\begin{figure}[!t]
 \centering
\begin{tabular}{cc}
\hspace*{-3ex}
\includegraphics[width=4.9cm,height=4.9cm]{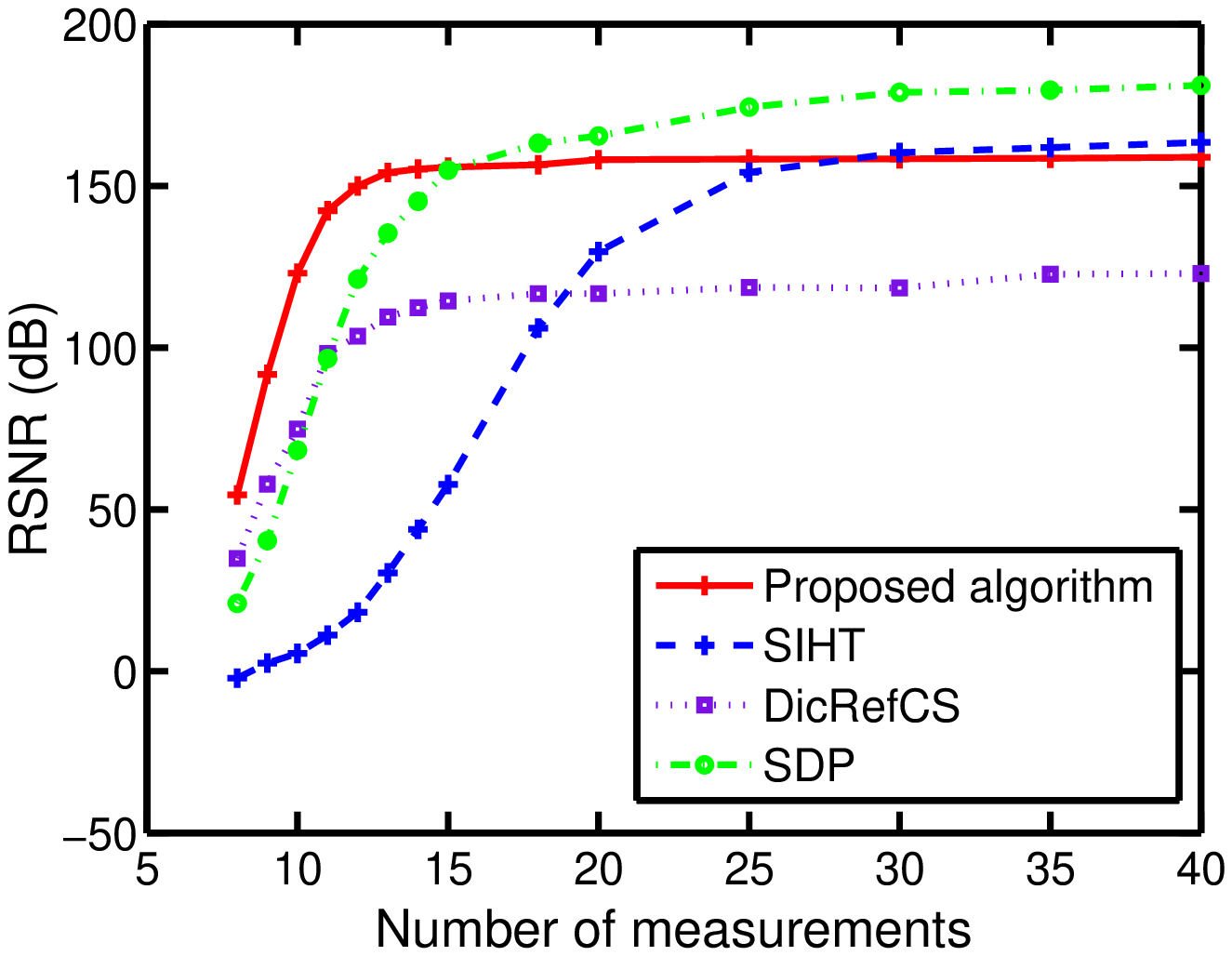}&
\hspace*{-5ex}
\includegraphics[width=4.9cm,height=4.9cm]{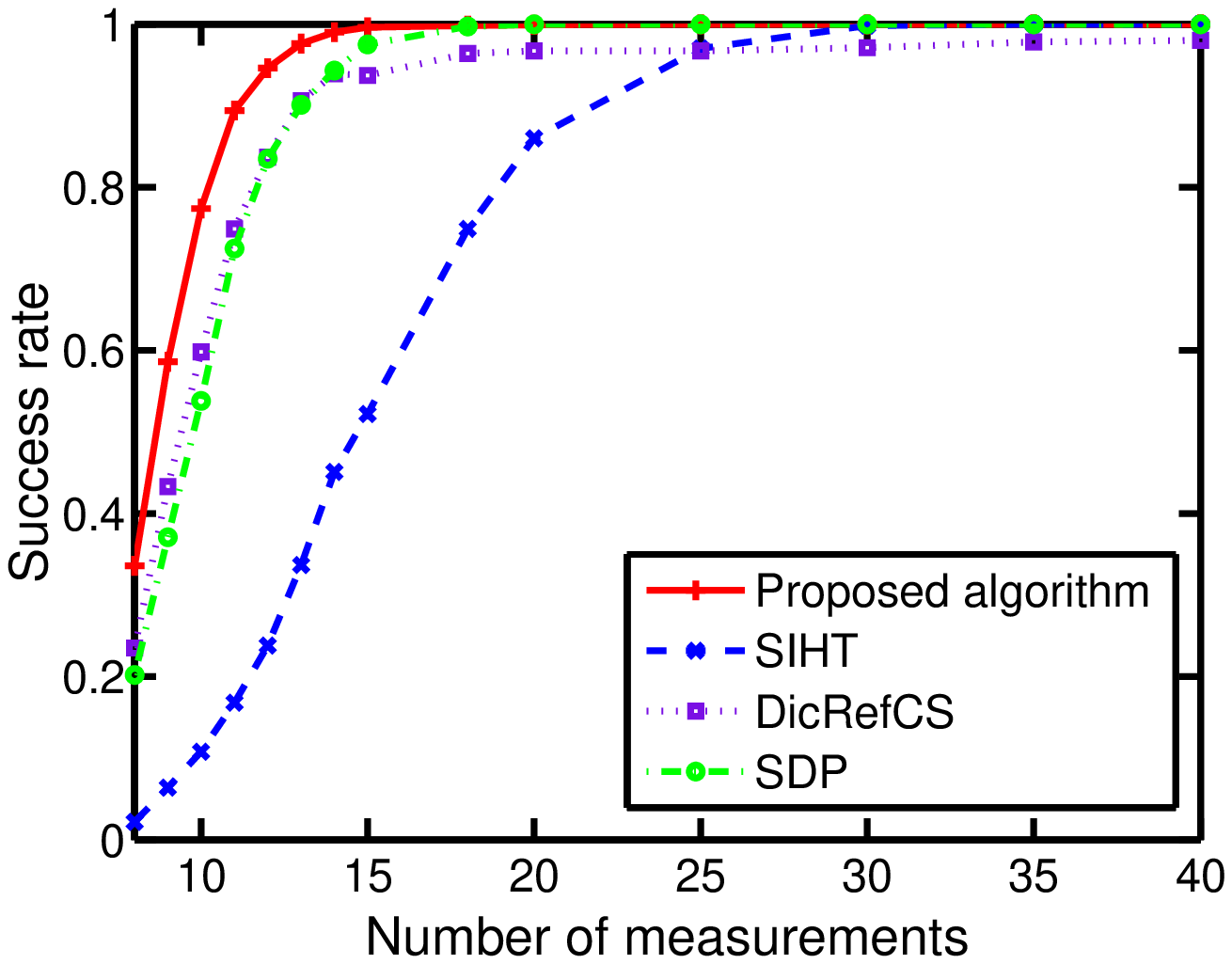}
\\
(a)& (b)
\end{tabular}
  \caption{(a). RSNRs of respective algorithms vs. $M$; (b). Success rates of respective algorithms vs. $M$.}
   \label{fig1}
\end{figure}

\begin{figure}[!t]
 \centering
\begin{tabular}{cc}
\hspace*{-3ex}
\includegraphics[width=4.9cm,height=4.9cm]{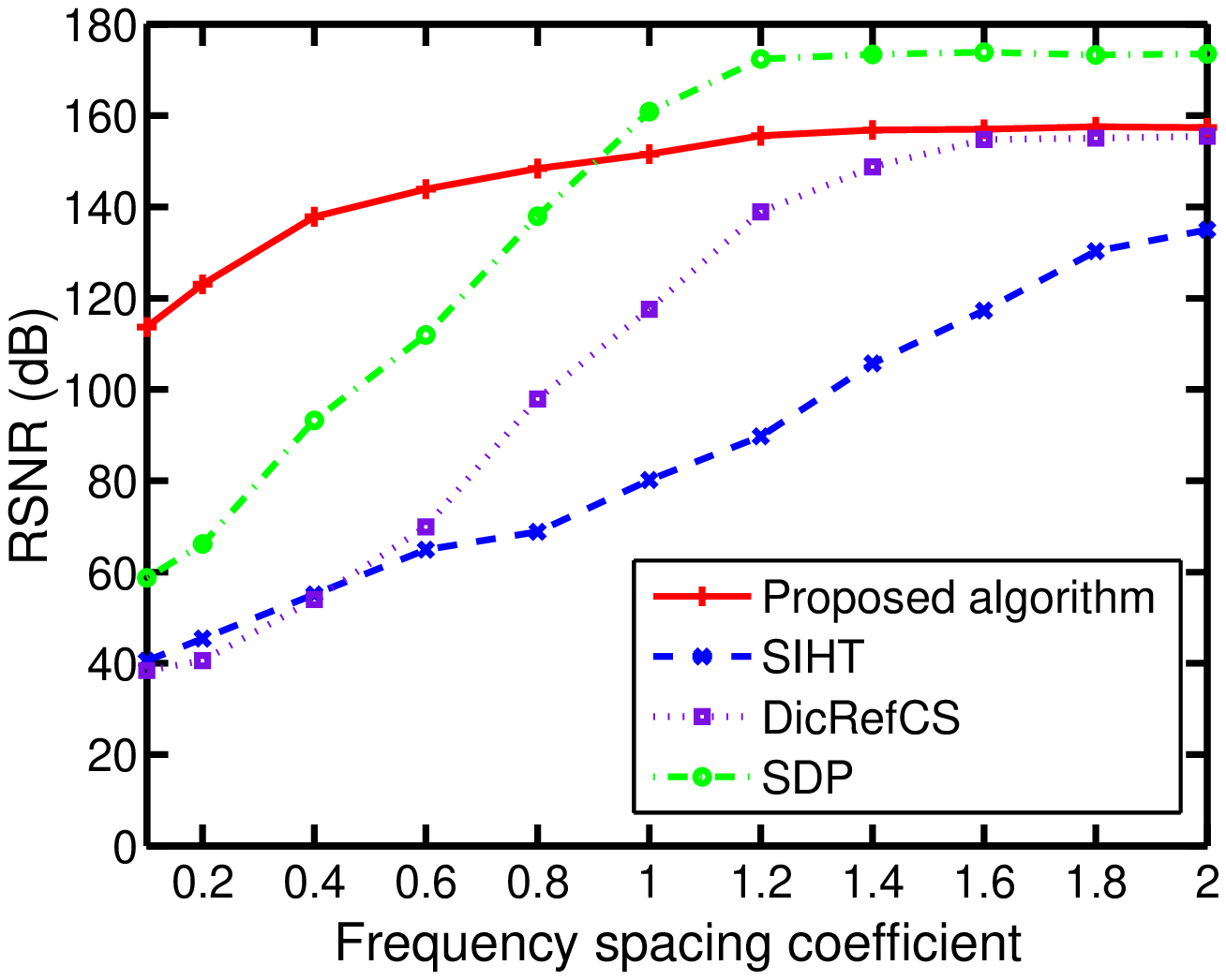}&
\hspace*{-5ex}
\includegraphics[width=4.9cm,height=4.9cm]{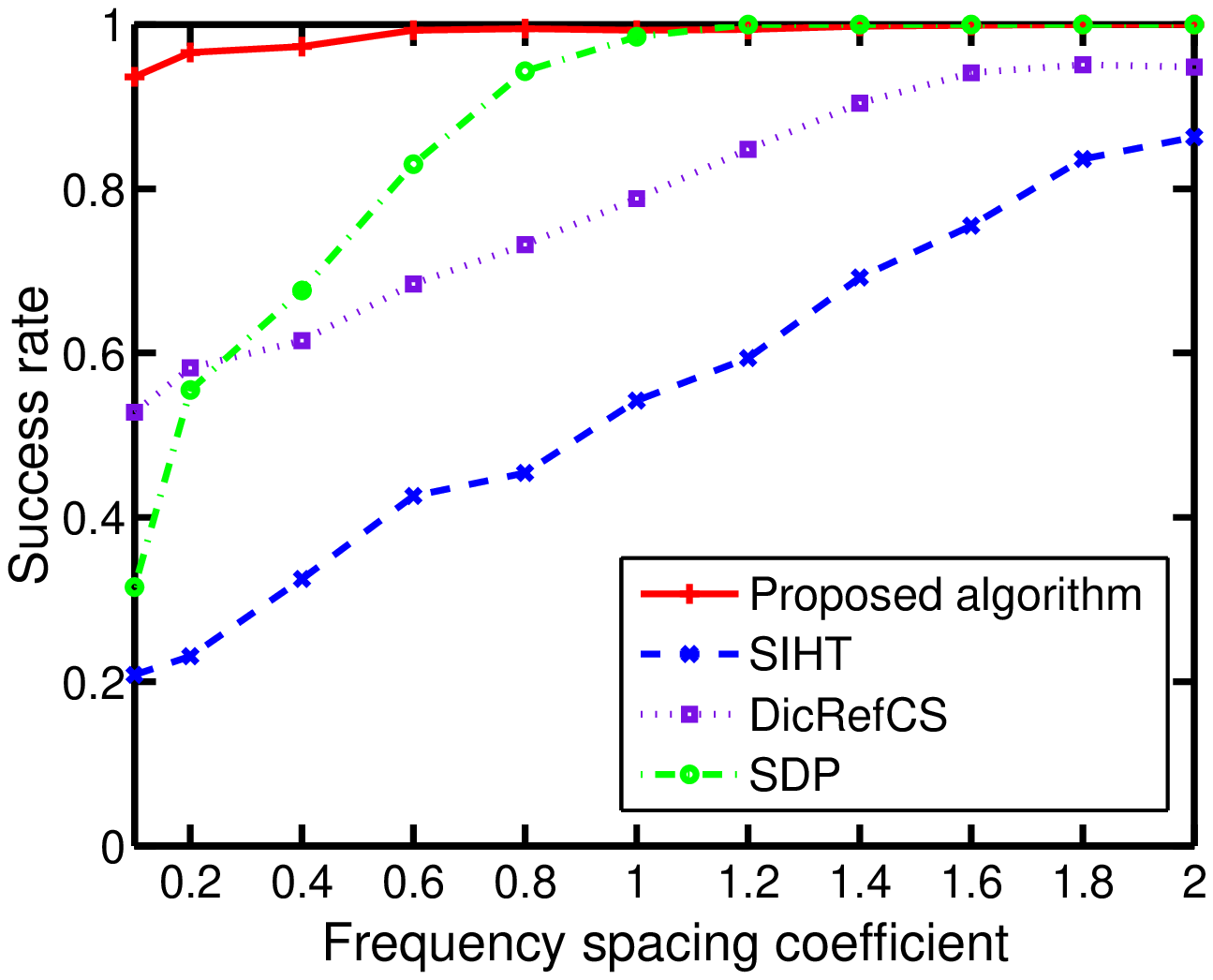}
\\
(a)& (b)
\end{tabular}
  \caption{(a). RSNRs of respective algorithms vs. the frequency spacing
coefficient $\mu$;
   (b). Success rates of respective algorithms vs. $\mu$.}
   \label{fig2}
\end{figure}

%\begin{figure}[!t]
% \centering
%\begin{tabular}{cc}
%\hspace*{-3ex}
%\includegraphics[width=8cm]{rsnr}&
%\hspace*{-5ex}
%\includegraphics[width=8cm]{success-rate}
%\\
%(a)& (b)
%\end{tabular}
%  \caption{(a). RSNRs of respective algorithms vs. $M$; (b). Success rates of respective algorithms vs. $M$.}
%   \label{fig1}
%\end{figure}

%\begin{figure}[!t]
% \centering
%\begin{tabular}{cc}
%\hspace*{-3ex}
%\includegraphics[width=8cm]{rsnr2}&
%\hspace*{-5ex}
%\includegraphics[width=8cm]{success-rate2}
%\\
%(a)& (b)
%\end{tabular}
%  \caption{(a). RSNRs of respective algorithms vs. the frequency spacing
%coefficient $\mu$;
%   (b). Success rates of respective algorithms vs. $\mu$.}
%   \label{fig2}
%\end{figure}

%also referred to as the iterative reweighted joint parameter
%learning and sparse signal recovery (IR-JPLSSR)

%of reconstructing a mixture of complex sinusoids from a small
%number of measurements.

\section{Simulation Results}
We now carry out experiments to illustrate the performance of our
proposed algorithm\footnote{Matlab codes are available at
http://www.junfang-uestc.net/codes/SRCS.rar} and its comparison
with other existing methods. We assume that the signal
$\boldsymbol{u}\triangleq [u_1\phantom{0}\ldots\phantom{0}u_L]^T$
is a mixture of $K$ complex sinusoids, i.e.
\begin{align}
u_l=\sum_{k=1}^K \alpha_k e^{-j\omega_k l} \qquad l=1,\ldots, L
\nonumber
\end{align}
with the frequencies $\{\omega_k\}$ uniformly generated over
$[0,2\pi)$ and the amplitudes $\{\alpha_k\}$ uniformly distributed
on the unit circle. The measurements $\boldsymbol{y}$ are obtained
by randomly selecting $M$ entries from $L$ elements of
$\boldsymbol{u}$. We first consider recovering the original signal
$\boldsymbol{u}$ from the partial observations $\boldsymbol{y}$.
The reconstruction accuracy is measured by the ``reconstruction
signal-to-noise ratio'' (RSNR) which is defined as
\begin{align}
\text{RSNR}=20\log_{10}\left(\frac{\|\boldsymbol{u}\|_2}{\|\boldsymbol{u}-\boldsymbol{\hat{u}}\|_2}\right)
\nonumber
\end{align}
We compare our proposed algorithm with the Bayesian dictionary
refinement compressed sensing algorithm (denoted as DicRefCS)
\cite{HuShi12}, the root-MUSIC based spectral iterative hard
thresholding (SIHT) \cite{DuarteBaraniuk13}, and the atomic norm
minimization via the semi-definite programming (SDP) approach
\cite{TangBhaskar12}. Fig. \ref{fig1}(a) depicts the average RSNRs
of respective algorithms as a function of the number of
measurements, $M$, where we set $L=64$ and $K=3$. Results are
averaged over $10^3$ independent runs, where the frequencies and
the sampling indices (used to obtain $\boldsymbol{y}$) are
randomly generated for each run. We observe that our proposed
algorithm outperforms the other three methods in the region of a
small $M$, where a gain of more than 15dB is achieved as compared
with the DicRefCS and SDP methods. Our algorithm is surpassed by
the SIHT and SDP methods as $M$ increases. Nevertheless, this
performance improvement is of less significance since all methods
provide quite decent recovery performance when $M$ is large.

%where data acquisition is more beneficial due to high compression
%rates
%it performs not as good as the other three algorithms when a
%limited number of measurements are available.

The recovery performance is also evaluated in terms of the success
rate. The success rate is computed as the ratio of the number of
successful trials to the total number of independent runs, where
$\{\alpha_k\}$ and $\{\omega_k\}$ are randomly generated for each
run. Note that our algorithm and the DicRefCS method do not
require the knowledge of the number of complex sinusoids, $K$. A
trial is considered successful if the number of frequency
components is estimated correctly\footnote{For our algorithm, some
of the coefficients of the estimated signal keep decreasing each
iteration, but will not exactly equal to zero. We assume that a
frequency is identified if the coefficient is greater than
$10^{-3}$.} and the estimation error between the estimated
frequencies $\{\hat{\omega}_k\}$ and the true parameters
$\{\omega_k\}$ is smaller than $10^{-3}$, i.e.
$\frac{1}{2\pi}\|\boldsymbol{\omega}-\boldsymbol{\hat{\omega}}\|_2\leq
10^{-3}$. Fig. \ref{fig1}(b) depicts the success rates of
respective algorithms vs. the number of measurements. This result
again demonstrates the superiority of the proposed algorithm over
other existing methods, particularly for the case when $M$ is
small.

We examine the ability of our algorithm in resolving
closely-spaced frequency components. The signal $\boldsymbol{u}$
is assumed a mixture of two complex sinusoids with the frequency
spacing $d_f\triangleq\frac{1}{2\pi}(\omega_1-\omega_2)$ equal to
$\mu/L$, where $\mu$ is the frequency spacing coefficient ranging
from $0.1$ to $2$. Fig. \ref{fig2} shows RSNRs and success rates
of respective algorithms vs. the frequency spacing coefficient
$\mu$, where we set $L=64$ and $M=20$. Results are averaged over
$10^3$ independent runs, with one of the two frequencies (the
other frequency is determined by the frequency spacing) and the
set of sampling indices randomly generated for each run. We see
that our algorithm can accurately identify closely-spaced (say,
$d_f=0.1/L$) frequencies with a high success rate and presents a
significant performance advantage over other methods when two
frequencies are very closely separated.

%We see that our proposed algorithm achieves the highest success
%rate among all methods.

%is able to estimate the unknown parameters associated with the
%sparsifying dictionary with a high precision

\section{Conclusions}
We proposed an iterative reweighted algorithm for joint parametric
dictionary learning and sparse signal recovery. The proposed
algorithm was developed by iteratively decreasing a surrogate
function majorizing the original objective function. Simulation
results show that the proposed algorithm presents superiority over
other existing methods in resolving closely-spaced frequency
components.

%and the basis pursuit method which uses an oversampled discrete
%Fourier transform (DFT) matrix with evenly-spaced frequency grid
%points as the presumed sparsifying dictionary

%For the DOA estimation problem, we use the success rate to
%evaluate the recovery performance.

%We now consider a DOA estimation problem where $K$ narrowband
%far-field sources $\{s_k\}$ impinging on a uniform linear array of
%$M$ sensors from directions $\{\theta_k\}$. The received signal
%can be expressed as
%\begin{align}
%\boldsymbol{y}(t)=\boldsymbol{A}(\boldsymbol{\theta})\boldsymbol{s}(t)+\boldsymbol{w}(t),
%\quad t=1\ldots, T \nonumber
%\end{align}
%where $\boldsymbol{w}(t)$ denotes the additive noise,
%$\boldsymbol{A}(\boldsymbol{\theta})=[\boldsymbol{a}(\theta_1)\phantom{0}\ldots\phantom{0}\boldsymbol{a}(\theta_K)]$
%is the array manifold matrix and $\boldsymbol{a}(\theta_k)$ is
%known as the steering vector of the $k$th source. In our
%simulations, we assume a noiseless case and there is only one
%snapshot of observation, i.e. $T=1$. The impinging directions
%$\{\theta_k\}$ are uniformly generated over $[0,\pi)$, and the
%signals $\{s_k\}$ are uniformly distributed on the unit circle.

\bibliography{newbib}

\begin{thebibliography}{10}
\providecommand{\url}[1]{#1}
\csname url@rmstyle\endcsname
\providecommand{\newblock}{\relax}
\providecommand{\bibinfo}[2]{#2}
\providecommand\BIBentrySTDinterwordspacing{\spaceskip=0pt\relax}
\providecommand\BIBentryALTinterwordstretchfactor{4}
\providecommand\BIBentryALTinterwordspacing{\spaceskip=\fontdimen2\font plus
\BIBentryALTinterwordstretchfactor\fontdimen3\font minus
  \fontdimen4\font\relax}
\providecommand\BIBforeignlanguage[2]{{%
\expandafter\ifx\csname l@#1\endcsname\relax
\typeout{** WARNING: IEEEtran.bst: No hyphenation pattern has been}%
\typeout{** loaded for the language `#1'. Using the pattern for}%
\typeout{** the default language instead.}%
\else
\language=\csname l@#1\endcsname
\fi
#2}}

\bibitem{CandesGranda12}
E.~Cand{\`e}s and C.~Fernandez-Granda, ``Towards a mathematical theory of
  super-resolution,'' \emph{Communications on Pure and Applied Mathematics}, to
  appear.

\bibitem{TangBhaskar12}
G.~Tang, B.~N. Bhaskar, B.~Recht, and P.~Shah, ``Compressed sensing off the
  grid,'' Available at http://arxiv.org/abs/1207.6053, 2012.

\bibitem{ChiScharf11}
Y.~Chi, L.~L. Scharf, A.~Pezeshki, and A.~R. Calderbank, ``Sensitivity to basis
  mismatch in compressed sensing,'' \emph{IEEE Trans. Signal Processing},
  vol.~59, no.~5, pp. 2182--2195, May 2011.

\bibitem{YangXie13}
Z.~Yang, L.~Xie, and C.~Zhang, ``Off-grid direction of arrival estimation using
  sparse {B}ayesian inference,'' \emph{IEEE Trans. Signal Processing}, vol.~61,
  no.~1, pp. 38--42, Jan. 2013.

\bibitem{HuZhou13}
L.~Hu, J.~Zhou, Z.~Shi, and Q.~Fu, ``A fast and accurate reconstruction
  algorithm for compressed sensing of complex sinusoids,'' \emph{IEEE Trans.
  Signal Processing}, to appear.

\bibitem{FannjiangLiao12}
A.~Fannjiang and W.~Liao, ``Coherence pattern-guided compressive sensing with
  unresolved grids,'' \emph{SIAM J. Imaging Sciences}, vol.~5, no.~1, pp.
  179--202, 2012.

\bibitem{DuarteBaraniuk13}
M.~F. Duarte and R.~G. Baraniuk, ``Spectral compressive sensing,''
  \emph{Applied and Computational Harmonic Analysis}, vol.~35, pp. 111--129,
  2013.

\bibitem{HuShi12}
L.~Hu, Z.~Shi, J.~Zhou, and Q.~Fu, ``Compressed sensing of complex sinusoids:
  An approach based on dictionary refinement,'' \emph{IEEE Trans. Signal
  Processing}, vol.~60, no.~7, pp. 3809--3822, 2012.

\bibitem{CoifmanWickerhauser92}
R.~R. Coifman and M.~Wickerhauser, ``Entropy-based algorithms for best basis
  selction,'' \emph{IEEE Trans. Information Theory}, vol. IT-38, pp. 713--718,
  Mar. 1992.

\bibitem{GorodnitskyRao97}
I.~F. Gorodnitsky and B.~D. Rao, ``Sparse signal reconstructions from limited
  data using focuss: A re-weighted minimum norm algorithm,'' \emph{IEEE Trans.
  Signal Processing}, vol.~45, no.~3, pp. 699--616, Mar. 1997.

\bibitem{CandesWakin08}
E.~Cand{\`e}s, M.~Wakin, and S.~Boyd, ``Enhancing sparsity by reweighted $l_1$
  minimization,'' \emph{Journal of Fourier Analysis and Applications}, vol.~14,
  pp. 877--905, Dec. 2008.

\bibitem{ChartrandYin08}
R.~Chartrand and W.~Yin, ``Iterative reweighted algorithm for compressive
  sensing,'' in \emph{IEEE International Conference on Acoustics, Speech, and
  Signal Processing}, Las Vegas, Nevada, USA, 2008.

\bibitem{ShenFang13}
Y.~Shen, J.~Fang, and H.~Li, ``Exact reconstruction analysis of log-sum
  minimization for compressed sensing,'' \emph{IEEE Signal Processing Letters},
  vol.~20, pp. 1223--1226, Dec. 2013.

\bibitem{AtaeeZayyani10}
M.~Ataee, H.~Zayyani, M.~Babaie-Zadeh, and C.~Jutten, ``Parametric dictionary
  learning using steepest descent,'' in \emph{IEEE International Conference on
  Acoustics, Speech, and Signal Processing, Proceedings}, Dallas, Texas, USA,
  2010.

\bibitem{LangeHunter00}
K.~Lange, D.~Hunter, and I.~Yang, ``Optimization transfer using surrogate
  objective functions,'' \emph{Journal of Computational and Graphical
  Statistics}, vol.~9, no.~1, pp. 1--20, Mar. 2000.

\end{thebibliography}
\bibliographystyle{IEEEtran}

\end{document}